\documentclass[11pt,a4paper]{article}
\usepackage[utf8]{inputenc}
\usepackage[english]{babel}
\usepackage[T1]{fontenc}
\usepackage{amsmath}
\usepackage{amssymb}
\usepackage{algorithm}
\usepackage[noend]{algpseudocode}
\usepackage[all,cmtip,curve]{xy}

\usepackage{paralist}
\usepackage{hyperref}

\def\land{\mathrel{\wedge}}

\algblockdefx{Payload}{EndPayload}[1]{\textbf{payload} #1}{end}
\algblockdefx{Query}{EndQuery}[3]{\textbf{query} \textsf{#1} (#2) : #3}{end}
\algblockdefx{Update}{EndUpdate}[2]{\textbf{update} \textsf{#1} (#2)}{end}
\algblockdefx{UpdateReturn}{EndUpdateReturn}[4]{\textbf{update} \textsf{#1} (#2) : #3 \emph{#4}}{end}
\algblockdefx{Merge}{EndMerge}[3]{\textbf{merge} (\emph{#1}, \emph{#2}) : payload \emph{#3}}{end}
\algblockdefx{Compare}{EndCompare}[3]{\textbf{compare} (\emph{#1}, \emph{#2}) : boolean \emph{#3}}{end}
\algblockdefx{UpdateOp}{EndUpdateOp}[1]{\textbf{update} \textsf{#1}}{end}
\algblockdefx{AtSource}{EndAtSource}[1]{\textbf{generator} (#1)}{end}
\algblockdefx{AtSourceReturn}{EndAtSourceReturn}[3]{\textbf{generator} (#1) : #2 \emph{#3}}{end}
\algblockdefx{AtSourceMultipleReturn}{EndAtSourceMultipleReturn}[2]{\textbf{generator} (#1) : #2}{end}
\algblockdefx{Downstream}{EndDownstream}[1]{\textbf{effector} (#1)}{end}
\newcommand{\Let}{\textbf{let} }
\algnewcommand{\IfThenElse}[3]{\State \algorithmicif\ #1\ \algorithmicthen\ #2\ \algorithmicelse\ #3}

\algnotext{EndPayload}
\algnotext{EndQuery}
\algnotext{EndUpdate}
\algnotext{EndUpdateOp}
\algnotext{EndUpdateReturn}
\algnotext{EndMerge}
\algnotext{EndCompare}
\algnotext{EndAtSource}
\algnotext{EndAtSourceReturn}
\algnotext{EndAtSourceMultipleReturn}
\algnotext{EndDownstream}

\newcommand{\note}[1]{}

\begin{document}

%-------------------------------------------------------------
\date{}

\title{Conflict-free Replicated Data Types: An Overview
%\textbf{**** Draft: Please do not Distribute ****}
}
\author{Nuno Preguiça\\
NOVA LINCS \& DI, FCT, Universidade NOVA de Lisboa}
\maketitle

\section{Introduction}

Internet-scale distributed systems often replicate data at multiple geographic locations
to provide low latency and high availability, despite node and network failures. 
Some systems \cite{spanner,replicatedcommit,farm,mdcc,Zhang13Transaction,blotter} adopt strong consistency models,
where the execution of an operation needs to involve coordination of a quorum of replicas.
Although it is possible to improve the throughput of these systems
(e.g., through batching), the required coordination leads to high latency for executing one operation,
that depends on the round-trip time among replicas and the protocol used.
Additionally, in the presence of a network partition or other faults, some nodes
might be unable to contact the necessary quorum for executing their operations.

An alternative approach is to rely on weaker consistency models, such as 
eventual consistency \cite{dynamo,lotusnotes,cassandra} or causal 
consistency \cite{cops,chainreaction,swiftcloud}, where any replica can 
accept updates, which are propagated asynchronously to other replicas. 
These models are also interesting for supporting applications running in mobile devices, 
to mask the high latency of mobile communications and the period of disconnection or poor
connectivity.

Systems that adopt a weak consistency model allow replicas to temporarily diverge, 
requiring a mechanism for merging concurrent updates into a common state. 
Conflict-free Replicated Data Types (CRDT) provide a principled approach to address this problem.

A CRDT is an abstract data type, with a well defined interface,
designed to be replicated at multiple nodes and exhibiting the following properties:
\begin{inparaenum}[(i)]
\item any replica can be modified without coordinating with any other replicas;
\item when any two replicas have received the same set of updates, 
they reach the same state, deterministically, 
by adopting mathematically sound rules to guarantee state convergence. 
\end{inparaenum}

Since our first works proposing a CRDT for concurrent 
editing \cite{DBLP:journals/corr/abs-0710-1784,Preguica09Commutative} and later laying the theoretical
foundations of CRDTs \cite{Shapiro11Conflict}, CRDTs have become mainstream 
and are used in a large number of systems serving millions of users
worldwide.
Currently, an application can use CRDTs by either using a storage system that offers
CRDTs in its interface,%\footnote{Examples of storage systems that offer CRDTs in their APIs include 
%Riak distributed database (\url{http://basho.com/products/riak-kv/}), 
%Redis CRDBs (\url{https://redislabs.com/redis-enterprise-documentation/developing/crdbs/}) and
%Akka distributed data (\url{https://doc.akka.io/docs/akka/current/distributed-data.html}).}
by embedding an existing CRDT library or implementing its own support.
 
This document presents an overview of Conflict-free Replicated Data Types
research and practice, organized as follows.

Section \ref{sec:appdev} discusses the aspects that are important 
for an \emph{application developer} that uses CRDTs to maintain the state
of her application.
As any abstract data type, a CRDT implements some given functionality and
important aspects that must be considered include 
the time complexity for operation execution and
the space complexity for storing data and for synchronizing replicas. 
However, as CRDTs are designed to be replicated and to allow uncoordinated 
updates, a key aspect of a CRDT is its semantics in the presence of concurrency
-- this section focuses on this aspect.

Section \ref{sec:sysdev} discusses the aspects that are important for the 
\emph{system developer} that needs to create a system that includes CRDTs.
These developers need to focus on another key aspect of CRDTs: the synchronization model. 
The synchronization model defines the requirements that the system must meet so that
CRDTs work correctly. 

Finally, Section \ref{sec:crdtdev} discusses the aspects that are important for the 
\emph{CRDT developer}, focusing on the key ideas and techniques used in 
the design of existing CRDTs.

\section{CRDTs for the Application Developer}\label{sec:appdev}

When developing an application, the application developer typically needs to decide 
how to maintain the application state. 
To this end, a common approach is to define the application data model and select 
the most appropriate data types for implementing this data model, given the 
functionality provided by the data types. 
CRDTs are data types designed to be modified concurrently. 
Thus, a first observation is that it only makes sense to adopt CRDTs if the 
application data can be modified concurrently.

An important aspect regarding the functionality of CRDTs is the 
concurrency semantics, that defines how the CRDT behaves when concurrent updates
are executed in different replicas, and later all updates are propagated 
to all replicas. 
In this section, we focus on this aspect, including an overview of the most 
relevant concurrency semantics proposed in literature for different data types.
We also discuss the general properties related with the concurrency semantics and
some advanced topics.

Another important aspect when selecting a concrete implementation of an abstract 
data type (ADT) is its performance, namely the complexity of the state representation and
operation execution. With CRDTs, besides this aspects, it is also important to 
consider the size of the data that is necessary for synchronizing replicas.
We defer the discussion of these aspects until Section~\ref{sec:crdtdev}, as
they are closely related with the implementation of CRDTs.

%==================================================================================================
%==================================================================================================
%==================================================================================================
\subsection{Concurrency semantics}\label{sec:appdev:conc_sem}

An abstract data type (or simply data type) defines a set of operations,
that can be classified in queries, when they have no influence in the
result of subsequent operations, and updates, when their execution may 
influence the result of subsequent operations.
In an implementation of a data type, a query will not modify the internal state
of the implementation, while an update might modify the internal state.

For the replication of an object, we consider a system with $n$ nodes.
Each node keeps a replica of the object. Applications interact with the an object
by executing operation in a replica of the object. Updates execute initially in a
replica and are propagated asynchronously to all other replicas. 
In Section \ref{sec:sysdev}, we discuss how updates are propagated.

The updates defined in a data type may intrinsically commute or not. 
Consider for instance a Counter, a shared integer that supports increments 
and decrements. 
As these updates commute (i.e., executing them in any order yields the same result), 
the Counter naturally converges towards the same expected result
independently of the order in which updates are applied.
In this case, it is natural that the state of a CRDT object reflects all 
executed updates.

Unfortunately, for most data types, this is not the case and several 
concurrency semantics are reasonable, with different semantics being 
suitable for different applications.
For instance, consider a shared set object supporting add and
remove updates. 
There is no correct outcome when concurrently adding and removing the same
element.

%For instance, consider a shared memory cell supporting the assignment 
%operation. 
%If the initial value is 0, the correct outcome for concurrently 
%assigning 1 and 2 is not well defined.

\paragraph{Happens-before relation:}
When defining the concurrency semantics, an important concept 
is that of the \emph{happens-before} relation \cite{Lamport78Time}. 
In a distributed system, an event $e_1$ \emph{happened-before} an event $e_2$, 
$e_1 \prec e_2$, iff:
\begin{inparaenum}[(i)]
\item $e_1$ occurred before $e_2$ in the same process; or 
\item $e_1$ is the event of sending message $m$, and $e_2$ is the event
of receiving that message; or
\item there exists an event $e$ such that $e_1 \prec e$ and $e \prec e_2$. 
\end{inparaenum}
When applied to CRDTs, we can say that an
update $u_1$ \emph{happened-before} an update $u_2$, $u_1 \prec u2$, iff the 
effects of $u_1$ had been applied in the replica where $u_2$ was executed 
initially.

As an example, if an event is ``\emph{Alice reserved the meeting room}'', it 
is relevant to know if that was known when ``\emph{Bob reserved the meeting room}''.
If that is the case, one reasonable semantics is to give priority to Alice's prior reservation.
Otherwise, the events were concurrent and the concurrency semantics must define 
some arbitration rule to give priority to one update over the other.
As discussed later, many CRDTs implements concurrency semantics that give priority 
to one update over the other concurrent updates.

\paragraph{Total order among updates:}
Another relation that can be useful for defining the concurrency semantics is 
that of a total order among updates and particularly a total order that 
approximates wall-clock time. 
In distributed systems, it is common to maintain nodes with their physical clocks 
loosely synchronized. 
When combining the clock time with a site identifier, we have unique 
timestamps that are totally ordered.
Due to the clock skew among multiple nodes, although these timestamps
approximate an ideal global physical time, they do not necessarily respect the happens-before
relation.
This can be achieved by combining physical and logical clocks, as
shown by Hybrid Logical Clocks \cite{DBLP:conf/opodis/KulkarniDMAL14}.

This relation allows to define the \emph{last-writer-wins} semantics, where the 
value written by the last writer wins over the values written previously, 
according to the defined total order.

We now show how these relations can be used to define sensible concurrency semantics
for CRDTs.
During our presentation, when defining the value of a CRDT and 
following Burckhardt et. al. \cite{Burckhardt14Replicated}\footnote{Zeller et. al. \cite{Zeller14Formal}
concurrently proposed a similar approach for specifying the value of a CRDT.}, we 
consider the value defined as a function of the set of updates $O$ known at 
a given replica, the happens-before relation, $\prec$, established among updates and, for some
data types, of the total order, $<$, defined among updates.

%==================================================================================================
%==================================================================================================
\subsubsection{Register}\label{sec:appdev:conc_sem:register}
A \emph{register} maintains an opaque value and provides a single update
that writes an opaque value: $\mathsf{wr}(\mathit{value})$. 
Two concurrency semantics have been proposed leading to two different CRDTs:
the \emph{multi-value} register and the \emph{last-writer-wins} register \cite{Shapiro11Conflict}.

\paragraph{Multi-value register:}
In the \emph{multi-value} register CRDT, all concurrently written values are kept.
In this case, the read operation returns the set of concurrently written values.
Formally, the value of a multi-value register is 
defined as the multi-set:
$\{v \mid \mathsf{wr}(v) \in O \land \nexists \mathsf{wr}(u) \in O \cdot \mathsf{wr}(v) \prec \mathsf{wr}(u)\}$.

\paragraph{Last-write wins (LWW) register:}
In the \emph{last-writer-wins} register CRDT, priority is given to the last writer,
and only this value is kept, if any. 
Formally, the value of a LWW register can be defined 
as a set that is either empty or holds a single value:
$\{v \mid \mathsf{wr}(v) \in O \land \nexists \mathsf{wr}(u) \in O \cdot \mathsf{wr}(v) < \mathsf{wr}(u)\}$.

%==================================================================================================
%==================================================================================================
\subsubsection{Counter}
A \emph{counter} data type maintains an integer and can be modified 
by updates $\mathsf{inc}$ and $\mathsf{dec}$, 
to increase and decrease by one unit its value, respectively
(this can easily generalize to arbitrary amounts).
As mentioned previously, as updates intrinsically commute, the natural
concurrency semantics for a counter CRDT \cite{Shapiro11Conflict} is to have a final state that 
reflects the effects of all executed updates. 
More formally, the value of the counter can be computed by 
counting the number of increments and subtracting the number of decrements: 
$\left| \{ \mathsf{inc} \mid \mathsf{inc} \in O \} \right| - \left| \{ \mathsf{dec} \mid \mathsf{dec} \in O \} \right|$. 

\paragraph{Counter with write update:}
Now consider that we want to add a write update $\mathsf{wr}(n)$, to set
the counter to a given value. 
This opens two questions related with the concurrency semantics.
First, what should be the final state in the presence of 
two or more concurrent write updates. 
Building on the semantics defined for the register, the 
the \emph{last-writer-wins} semantics seems a good alternative 
(as maintaining multiple values, as in the multi-value register, 
seems rather complex and non intuitive in this case).

Second, what is the result when concurrent writes and $\mathsf{inc}$/$\mathsf{dec}$ updates 
are executed. 
In this case, by building on the happens-before relation, we can 
define several concurrency semantics.  
One possibility is a \emph{write-wins} semantics, where $\mathsf{inc}$/$\mathsf{dec}$ 
updates have no effect when executed concurrently with the 
last write.
Formally, for a given set $O$ of updates that include at least a write update \footnote{For making this definition general, one can consider that when a counter is created, a $\mathsf{wr}(0)$ is always executed.},
let $\mathsf{wr}(v)$ be the last write, i.e., $\mathsf{wr}(v) \in O \land \nexists \mathsf{wr}(u) \in O \cdot \mathsf{wr}(v) < \mathsf{wr}(u)$. 
The value of the counter would be
$v + o$, 
with $o = \left| \{ \mathsf{inc} \mid \mathsf{inc} \in O \wedge \mathsf{wr}(v) \prec \mathsf{inc}\} \right| - \left| \{ \mathsf{dec} \mid \mathsf{dec} \in O \wedge \mathsf{wr}(v) \prec \mathsf{dec}\} \right|$
the difference between the number of $\mathsf{inc}$ and $\mathsf{dec}$ updates that happened after the last write.

Other possibility is a \emph{inc/dec-write merge} semantics, where $\mathsf{inc}$/$\mathsf{dec}$ 
updates concurrent with the last write are considered for determining the value
of the counter.  
Formally, for a given set $O$ of updates that include at least a write update, 
let $\mathsf{wr}(v)$ be the last write, i.e., $\mathsf{wr}(v) \in O \land \nexists \mathsf{wr}(u) \in O \cdot \mathsf{wr}(v) < \mathsf{wr}(u)$. 
The value of the counter would be
$v + o$, 
with $o = \left| \{ \mathsf{inc} \mid \mathsf{inc} \in O \wedge \mathsf{inc} \not \prec \mathsf{wr}(v)\} \right| - \left| \{ \mathsf{dec} \mid \mathsf{dec} \in O \wedge \mathsf{dec} \not \prec \mathsf{wr}(v)\} \right|$
the difference between the number of $\mathsf{inc}$ and $\mathsf{dec}$ updates that have not happened before the last write.

%==================================================================================================
%==================================================================================================
\subsubsection{Set}

A \emph{set} data type provides two updates:
\begin{inparaenum}[(i)]
\item $\mathsf{add}(e)$, for adding element $e$ to the set; and 
\item $\mathsf{rmv}(e)$, for removing element $e$ from the set.
\end{inparaenum}
In the presence of a concurrent add and remove of the same element,
several concurrency semantics are possible.

\paragraph{Add-wins Set}
In the \emph{add-wins} semantics, intuitively, in the presence of concurrent 
add and remove updates, priority is given to add updates.
Thus, in the \emph{add-wins} set (also known as observed-remove set, OR-set 
\cite{Shapiro11Conflict}), in the presence of two updates that do not 
naturally commute, a concurrent add and remove of the same element, the 
add wins leading to a state where the element belongs to the set.

More formally, given a set of updates $O$, the elements of the set are:
$\{e \mid \mathsf{add}(e) \in O \land \nexists \mathsf{rmv}(e) \in O \cdot \mathsf{add}(e) \prec \mathsf{rmv}(e)\}$.

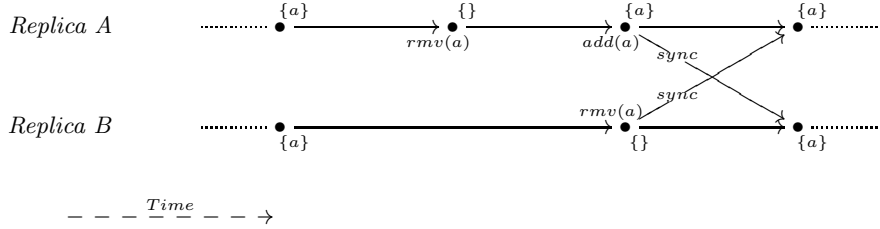
\begin{figure*}[h!]
\begin{center}
\footnotesize
\centerline{\begin{xy}
\xymatrix{
  \mathit{Replica\; A} & \ar@{.}[r] &
  {\bullet} \ar@{->}[rr]^<{\{a\}}_>{rmv(a)} & & 
  {\bullet} \ar@{->}[rr]^<{\{\}}_>{add(a)} & &
  {\bullet} \ar@{->}[rr]^<{\{a\}}\ar@{->}[rrd] |(0.3){sync} & & 
  {\bullet} \ar@{.}[r]^<{\{a\}} &\\
  \mathit{Replica\; B} & \ar@{.}[r] &
  {\bullet} \ar@{->}[rrrr]_<{\{a\}}^>{rmv(a)} & & & & 
  {\bullet} \ar@{->}[rr]_<{\{\}} \ar@{->}[rru] |(0.3){sync} & & 
  {\bullet} \ar@{.}[r]_<{\{a\}} & \\
    \ar@{-->}[rr] ^{Time} & & &  & & & & &  
}
\end{xy} }
\end{center}
\caption{Run with an add-wins set.}
\label{fig:set:add-wins}
\end{figure*}

Figure~\ref{fig:set:add-wins} shows a run where an add-wins set is replicated
in two replicas, with initial state $\{a\}$. 
In this example, in replica A, $a$ is first removed and later added again to 
the set. In replica B, $a$ is removed from the set.
After these operations, each replica propagates its new updates to the other replica 
-- \emph{sync} arrow (the way updates are propagated is discussed later, in 
Section \ref{sec:sysdev}).
After receiving the updates from the other replica, both replicas end up 
with element $a$ in the set. The reason for this is that there is no
$\mathsf{rmv}(a)$ that happened after the $\mathsf{add}(a)$ executed in replica A.

\paragraph{Remove-wins Set}
An alternative semantics is to give priority to removes.
Intuitively, in the \emph{remove-wins} semantics, 
in the presence of a concurrent add and remove of the same element, the 
remove wins leading to a state where the element is not in the set.

More formally, given a set of updates $O$, the elements of the set 
are:
$\{e \mid \mathsf{add}(e) \in O \land \forall \mathsf{rmv}(e) \in O \cdot  \mathsf{rmv}(e) \prec \mathsf{add}(e)\}$.
In the previous example, after receiving the updates from the other replica, 
the state of both replicas 
would be the empty set, because there is no $\mathsf{add}(a)$ that happened after the
$\mathsf{rmv}(a)$ in replica B.

\paragraph{Last-writer-wins (LWW) Set:}
A third concurrency semantics if to give priority to updates based 
on a total order defined among them, for example using the \emph{last-writer-wins}
semantics defined before. 
Intuitively, in a \emph{last-writer-wins} set, in the presence of a concurrent 
add and remove of the same element, the element will be in the set if the add 
is ordered after the remove in the total order among updates.

More formally, with the set $O$ of updates now totally ordered by $<$, the elements
of a LWW set are:
$\{e \mid \mathsf{add}(e) \in O \land \forall \mathsf{rmv}(e) \in O \cdot \mathsf{rmv}(e) < \mathsf{add}(e)\}$.
Returning to our previous example, the state of the replicas after the
synchronization would include $a$ if, according the total order defined 
among the updates, the $\mathsf{rmv}(a)$ of replica B is smaller than
the $\mathsf{add}(a)$ of replica A. Otherwise, the state would be the empty set.

%==================================================================================================
%==================================================================================================
\subsubsection{List / sequence}
A \emph{list} (or \emph{sequence}) data type maintains an ordered collection of 
elements, and exports two updates:
\begin{inparaenum}[(i)]
\item $\mathsf{ins}(i,e)$, for inserting element $e$ in the position $i$, shifting 
element in position $i$, if any, and subsequent elements to the right; and 
\item $\mathsf{rmv}(i)$, for removing element in the position $i$, if any, and shifting
subsequent elements to the left.
\end{inparaenum}

\begin{figure*}[h!]
\begin{center}
\footnotesize
\centerline{\begin{xy}
\xymatrix{
  \mathit{Replica\; A} & \ar@{.}[r] &
  {\bullet} \ar@{->}[rr]^<{012345}_>{ins(2,A)} & & 
  {\bullet} \ar@{->}[rr]^<{01A2345}\ar@{->}[rrd] |(0.3){sync} & & 
  {\bullet} \ar@{.}[r]^<{01A23B45} &\\
  \mathit{Replica\; B} & \ar@{.}[r] &
  {\bullet} \ar@{->}[rr]_<{012345}^>{ins(4,B)} & &
  {\bullet} \ar@{->}[rr]_<{0123B45} \ar@{->}[rru] |(0.3){sync} & & 
  {\bullet} \ar@{.}[r]_<{01A23B45} & \\
    \ar@{-->}[rr] ^{Time} & & &  & & & & &  
}
\end{xy} }
\end{center}
\caption{Run with a list.}
\label{fig:list}
\end{figure*}
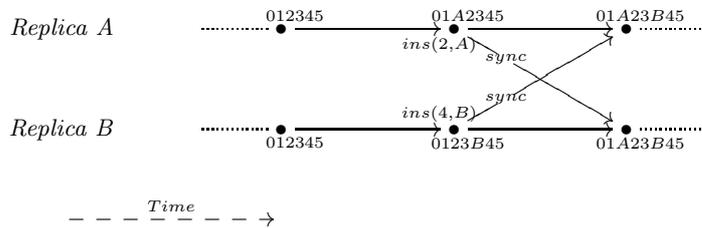

The problem in implementing a list CRDT \cite{Preguica09Commutative,Weiss09Logoot,Roh11Replicated} 
is that the position of elements
is shifted when inserting and deleting an element. For example, consider the example of 
Figure~\ref{fig:list}, where both replicas start with the list of six 
characters ``012345''.
In replica A, element A is inserted in position 2 (considering the first position of 
the list as having position 0).
In replica B, element B is inserted in position 4. 
When synchronizing, if updates would be replayed by executing the original operations, 
in replica A, B would be inserted before the ``3'' leading to ``01A2B345'', as this is 
the element in position 4 after inserting element ``A''. 

The concurrency semantics of lists have been studied extensively in the context of 
collaborative editing systems \cite{Sun98Operational}, with the following being 
the generally accepted correct semantics. 
A $\mathsf{rmv(i)}$ update should remove the element present in position $i$ in 
the replica where the update was initially executed.
An $\mathsf{ins(i,e)}$ update inserts element $e$ after the elements that precede 
element in position $i$ in the replica where the update was initially executed, 
and before the subsequent elements.
In the presence of concurrent updates, if the previous rule does not define a total
order on the elements, the order of the elements that could be in the same position
is arbitrated using some deterministic rule (that must guarantee that the relative 
order of elements remains constant over time) \footnote{This corresponds to the
strong list specification proposed by Attyia et. al. \cite{Attiya16Specification}. 
In the same paper, the authors also introduce a weak list specification, where the 
orderings relative to removed elements do not need to hold after their removal.
We conjecture that there is no CRDT design that allows peer-to-peer synchronization
and implements the weak list specification 
without implementing the strong list specification where the relative order of elements
remains constant over time. The rationale is that with peer-to-peer synchronization, 
when a remove executes concurrently with multiple inserts, there might a replica where
inserts arrive before the remove -- thus, the relative order of the inserted elements
cannot take into consideration the concurrently removed element. Consider the example
given in the paper, with the initial list value ``x''. On the concurrent execution of
$\mathsf{rmv(0)}$, $\mathsf{ins(0,'a')}$ and , $\mathsf{ins(1,'b')}$, the weak list 
specification allows the final result to be ``ba''. However, in a replica of a CRDT 
where the two insert updates arrive before the remove, the relative order must be
``a'' (before ``x'') before ``b'', which precludes the possibility of taking advantage 
of the ordering flexibility provided by the weak list specification.}.

Returning to our example, the final result in both replicas should be ``01A23B45'' because 
when $\mathsf{ins(4,B)}$ executed in replica B, it inserts ``B'' between 
elements ``3'' and ``4''. Thus, when the update is applied in replica A, it 
should also insert ``B'' between elements ``3'' and ``4''.

%==================================================================================================
%==================================================================================================
\subsubsection{Map}
We now consider a \emph{map} data type that maps keys to objects, exporting
two updates:
\begin{inparaenum}[(i)]
\item $\mathsf{put}(k,o)$, that associates key $k$ with object $o$; and  
\item $\mathsf{rmv}(k)$, that removes the mapping for key $k$, if any.
\end{inparaenum}

\paragraph{Map of literals:}
If a map can only contain literals, two questions must be answered
by the concurrency semantics.
First, what is the value associated with a key $k$ in the presence of
two concurrent puts for $k$. Building on the semantics of 
registers, it would be possible to adopt a \emph{last-writer-wins}
semantics, in which the value associated with a key is that of the 
latest put, or a \emph{multi-value} entry semantics, in which the map
would record the values of all concurrent put updates.

Second, it is necessary to decide what happens in the presence of a 
concurrent remove and put of the same key. 
By analogy with the set semantics, 
possible concurrency semantics include the \emph{put-wins}
semantics, the \emph{remove-wins} semantics and the \emph{last-write-wins}
semantics.

We can more formally define the entries of a map that combines the 
\emph{multi-value} entry semantics for handling concurrent puts and 
the \emph{remove-wins} semantics for handling concurrent remove and put updates
as follows:
$\{(k,v) \mid \mathsf{put}(k,v) \in O \land \forall \mathsf{rmv}(k) \in O \cdot \mathsf{rmv}(k) \prec \mathsf{put}(k,v) \land \nexists \mathsf{put}(k,v') \in O \cdot \mathsf{put}(k,v) \prec \mathsf{put}(k,v')\}$,
where a key $k$ has associated a value $v$ iff all $\mathsf{rmv}(k)$ updates happened-before
$\mathsf{put}(k,v)$ (remove-wins semantics for handling $\mathsf{put}$/$\mathsf{rmv}$ ) and
there is no $\mathsf{put}(k,v')$ that happened-after $\mathsf{put}(k,v)$ (leading to a multi-value
semantic for concurrent put, as multiple concurrent puts satisfy this condition).

\paragraph{Map of CRDTs:}
A more interesting case is to allow to associate a key with a CRDT
(we call such CRDT, an embedded CRDT\footnote{The semantics presented in 
this section could be applied also for embedding CRDTs in other \emph{container} 
CRDTs, such as a list or a set.}).
In this case, besides the put and remove updates, we must consider the fact
that some updates update the state of the embedded CRDT -- we refer
to these updates generically as $\mathsf{upd}(k,op)$. 
Using this formulation, we can consider that the put update, that associates a key
with an object, can be encoded as an update $\mathsf{upd}(k,init(o))$
that sets the initial value of the object associated with $k$.

The map allows embedding another map, leading to a recursive data type.
For simplicity of presentation, we consider 
that the key is simple, although when a map embeds another map, the 
key will be composed by multiple parts, one for each of the maps.

For defining the concurrency semantics of a map, it is necessary to consider 
the following cases.

First, the case of concurrent updates performed to the same embedded
object. In this case, with the objects being CRDTs, a natural choice is 
to rely on the semantics of the embedded CRDT to combine the 
concurrent updates.

An aspect that must be considered is that updates executed for a given key
should be compatible with the type of the object associated with the key.
A special case is when first associating an object with a key, it is possible
that objects of different types are associated with the same key.
A pragmatic solution to address this issue, proposed by Riak developers,
is to have, for each key, one object for each CRDT type \cite{Brown14Riak}.
In this case, when accessing the value of a key, it is necessary to specify 
which type should be accessed -- this can be seen as the key of the 
map is the pair \emph{(key, data type)}.

Second, the case of a concurrent remove of the key and update of the object 
associated with the key (or of an object embedded in the object associated
with the key).
To address this case, several concurrency semantics can be proposed.

\subparagraph{Remove-as-recursive-reset map:}
A first possible concurrency semantics is the \emph{remove-as-recursive-reset}, where 
a remove of a key $k$ is transformed in executing a reset update 
in the object $o$ associated with $k$, and recursively in all objects 
embedded in $o$. 
A reset update is a type-specific operation that sets the value of the
object to a bottom value.

Concurrent updates to the same object, including reset updates, are 
then solved by relying on the concurrency semantics defined for the CRDT.
This approach requires every object that can be embedded to define a reset 
update.
Additionally, in the concrete implementations 
of the designs proposed in literature \cite{Brown14Riak,DBLP:journals/jpdc/AlmeidaSB18}, 
for some data types, it might be difficult to have a bottom value that differs
from a value of the domain -- e.g. for counter, 0 is often used as the bottom value.
In this case, it might become impossible to distinguish between a removed object 
and an object that was assigned the bottom value.

\begin{figure*}[h!]
\begin{center}
\footnotesize
\centerline{\begin{xy}
\xymatrix{
  \mathit{Replica\; A} & \ar@{.}[r] &
  {\bullet} \ar@{->}[rr]^<{\txt{\tiny \{ sugar $\rightarrow$ 1\\\tiny flour $\rightarrow$ 2\}}}_>{upd(flour,inc)} & & 
  {\bullet} \ar@{->}[rrr]^<{\txt{\tiny \{ sugar $\rightarrow$ 1\\\tiny flour $\rightarrow$ 3\}}} & \ar@{->}[rrd] |(0.3){sync} & & 
  {\bullet} \ar@{.}[r]^<{\txt{\tiny \{ flour $\rightarrow$ 1\}}} &\\
  \mathit{Replica\; B} & \ar@{.}[r] &
  {\bullet} \ar@{->}[rr]_<{\txt{\tiny \{ sugar $\rightarrow$ 1\\\tiny flour $\rightarrow$ 2\}}}^>{\txt{\tiny rmv(sugar)\\\tiny rmv(flour)}} & &
  {\bullet} \ar@{->}[rrr]_<{\txt{\tiny \{ \}}} & \ar@{->}[rru] |(0.3){sync} & & 
  {\bullet} \ar@{.}[r]_<{\txt{\tiny \{ flour $\rightarrow$ 1\}}} & \\
    \ar@{-->}[rr] ^{Time} & & & & & & & & &  
}
\end{xy} }
\end{center}
\caption{Run with a \emph{remove as recursive reset} map.}
\label{fig:map:rrr}
\end{figure*}
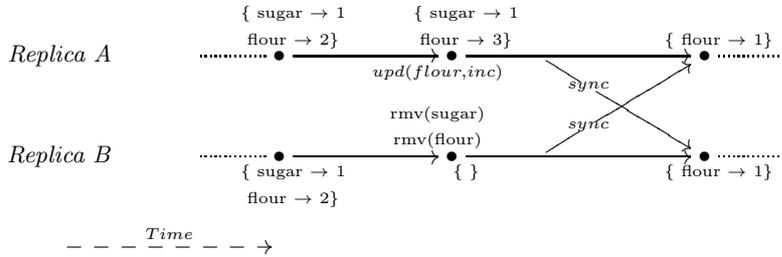

Consider the example of Figure~\ref{fig:map:rrr}, where the map is used to 
keep a shared shopping list, where a product is associated with a counter.
Now suppose that in replica A the entry ``flour'' is incremented (e.g. because one 
of the users of the shopping list realized that he needs more flour to bake a cake).
Concurrently, another user has done a checkout, which, as a side effect removed all
entries in the shopping list.
After synchronizing, the state of the map will show the value of 1 associated with 
``flour'', as the reset update for counters just sets the value to 0. 
This seems a sensible semantics for this use-case.

\subparagraph{Remove-wins map:}
A second possible concurrency semantics is the \emph{remove-wins} semantics, 
that gives priority to removes over updates. 
In this case, intuitively, a remove of key $k$ cancels the effects of all updates to $k$ 
(or any descendant of $k$) that either happened-before or are concurrent with the remove.

More formally, given the full set of updates $O$, the set of updates that must be 
considered for determining the final state of a map is 
$O' = \{\mathsf{upd}(k,op) \in O \mid \forall \mathsf{rmv}(k') \in O \cdot \mathit{prefix}(k',k) \Rightarrow 
\mathsf{rmv}(k') \prec \mathsf{upd}(k,op)\}$, i.e., 
all updates to a key $k$, such that for all $\mathsf{rmv}(k')$, with $k'$ a prefix of $k$,
the update happened after the remove of $k'$.

\begin{figure*}[h!]
\begin{center}
\footnotesize
\centerline{\begin{xy}
\xymatrix{
  \mathit{Replica\; A} & \ar@{.}[r] &
  {\bullet} \ar@{->}[rrr]^<{\txt{\tiny \{ Alice $\rightarrow$ \{ Coin $\rightarrow$ 10\\\tiny Objects $\rightarrow$ \{hammer\}\}\}}}_>{\txt{\tiny upd(Alice.Objects,\\\tiny add(nail))}} & & & 
  {\bullet} \ar@{->}[rrr]^<{\txt{\tiny \{ Alice $\rightarrow$ \{ Coin $\rightarrow$ 10\\\tiny Objects $\rightarrow$ \{hammer,nail\}\}\}}} & \ar@{->}[rrd] |(0.3){sync} & & 
  {\bullet} \ar@{.}[r]^<{\txt{\tiny \{ \}}} &\\
  \mathit{Replica\; B} & \ar@{.}[r] &
  {\bullet} \ar@{->}[rrr]_<{\txt{\tiny \{ Alice $\rightarrow$ \{ Coin $\rightarrow$ 10\\\tiny Objects $\rightarrow$ \{hammer\}\}\}}}^>{rmv(Alice)} & & &
  {\bullet} \ar@{->}[rrr]_<{\txt{\tiny \{ \}}} & \ar@{->}[rru] |(0.3){sync} & & 
  {\bullet} \ar@{.}[r]_<{\txt{\tiny \{ \}}} & \\
    \ar@{-->}[rr] ^{Time} & & & & & & & & & &  &
}
\end{xy} }
\end{center}
\caption{Run with a \emph{remove-wins} map.}
\label{fig:map:rmvwins}
\end{figure*}
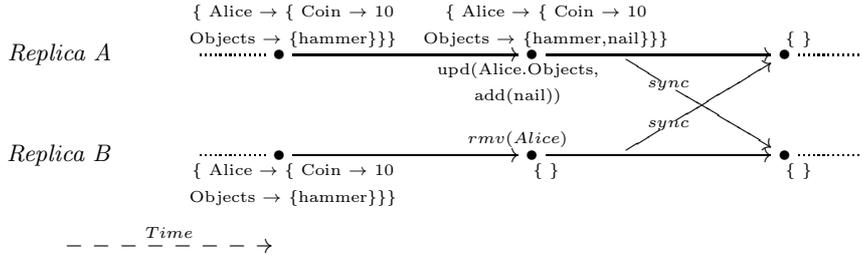

Consider the example of Figure~\ref{fig:map:rmvwins} where a map is used 
to store the state of a game. Player Alice has 10 coins and she has collected an hammer.
Now suppose that the system processes concurrently the following operations. 
In replica A, the state is updated by reflecting that Alice has collected a nail. 
In replica B, Alice is removed from the game, by removing the contents of her state.
Using the remove-wins semantics, when combining both updates, the state of the map
will include no information for Alice, as the remove wins over concurrent updates.

This is a sensible semantics, assuming that removing a player is a definite action and
we do not want to keep any state about the removed player.
We note that if we have used the remove as recursive reset semantics, the final state
would include for Alice, only the concurrently inserted object, the nail. 
This seems unreasonable in this use-case.

\subparagraph{Update-wins map:}

\begin{figure*}[h!]
\begin{center}
\footnotesize
\centerline{\begin{xy}
\xymatrix{
  \mathit{Replica\; A} & \ar@{.}[r] &
  {\bullet} \ar@{->}[rrr]^<{\txt{\tiny \{ Alice $\rightarrow$ \{ Coin $\rightarrow$ 10\\\tiny Objects $\rightarrow$ \{hammer\}\}\}}}_>{\txt{\tiny upd(Alice.Objects,\\\tiny add(nail))}} & & & 
  {\bullet} \ar@{->}[rrrr]^<{\txt{\tiny \{ Alice $\rightarrow$ \{ Coin $\rightarrow$ 10\\\tiny Objects $\rightarrow$ \{hammer,nail\}\}\}}} & \ar@{->}[rrd] |(0.3){sync} & & &
  {\bullet} \ar@{.}[r]^<{\txt{\tiny \{ Alice $\rightarrow$ \{ Coin $\rightarrow$ 10\\\tiny Objects $\rightarrow$ \{hammer,nail\}\}\}}} &\\
  \mathit{Replica\; B} & \ar@{.}[r] &
  {\bullet} \ar@{->}[rrr]_<{\txt{\tiny \{ Alice $\rightarrow$ \{ Coin $\rightarrow$ 10\\\tiny Objects $\rightarrow$ \{hammer\}\}\}}}^>{rmv(Alice)} & & &
  {\bullet} \ar@{->}[rrrr]_<{\txt{\tiny \{ \}}} & \ar@{->}[rru] |(0.3){sync} & & &
  {\bullet} \ar@{.}[r]_<{\txt{\tiny \{ Alice $\rightarrow$ \{ Coin $\rightarrow$ 10\\\tiny Objects $\rightarrow$ \{hammer,nail\}\}\}}} & \\
    \ar@{-->}[rr] ^{Time} & & & & & & & & & & &  &
}
\end{xy} }
\end{center}
\caption{Run with an \emph{update-wins} map.}
\label{fig:map:updwins}
\end{figure*}
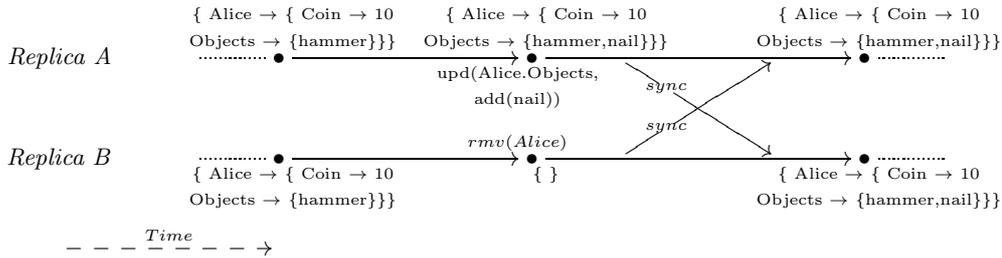

A third possible semantics is the \emph{update-wins} semantics, that gives
priority to updates over removes. 
Intuitively, we want an update to cancel the effects of a concurrent remove. 
In the previous example, we want the final state to reflect all updates that 
modified Alice's state, as adding a nail to her \emph{Objects} cancels the effect 
of the concurrent remove -- the expected run is presented in Figure~\ref{fig:map:updwins}.
This could be used, for example, in a situation where a player is removed because she 
does not execute updates for some period of time. 
In such case, if there is an update concurrent with the remove, it seems sensible
to restore the player's state and reflect all updates executed in that state. 
This would not be the case if any of the previous semantics was used.

\begin{figure*}[h!]
\begin{center}
\footnotesize
\centerline{\begin{xy}
\xymatrix{
  \mathit{Replica\; A} & \ar@{.}[r] &
  {\bullet} \ar@{->}[rr]^<{\txt{\tiny \{ Alice $\rightarrow$ \{ Coin $\rightarrow$ 10\\\tiny Objects $\rightarrow$ \{hammer\}\}\}}}_>{rmv(Alice)} & & 
  {\bullet} \ar@{->}[rr]^<{\txt{\tiny \{ \}}}_>{upd(Alice.Coin, wr(5))} & & 
  {\bullet} \ar@{->}[rrr]^<{\txt{\tiny \{ Alice $\rightarrow$ \{ Coin $\rightarrow$ 5\}\}}} & \ar@{->}[rrd] |(0.3){sync} & & 
  {\bullet} \ar@{.}[r]^<{\txt{\tiny \{ Alice $\rightarrow$ \{ Coin $\rightarrow$ 5\}\}}} & \\
  \mathit{Replica\; B} & \ar@{.}[r] &
  {\bullet} \ar@{->}[rr]_<{\txt{\tiny \{ Alice $\rightarrow$ \{ Coin $\rightarrow$ 10\\\tiny Objects $\rightarrow$ \{hammer\}\}\}}}^>{rmv(Alice)} & & 
  {\bullet} \ar@{->}[rr]_<{\txt{\tiny \{ \}}}^>{upd(Alice.Coin, wr(5))} & & 
  {\bullet} \ar@{->}[rrr]_<{\txt{\tiny \{ Alice $\rightarrow$ \{ Coin $\rightarrow$ 5\}\}}} & \ar@{->}[rru] |(0.3){sync} & & 
  {\bullet} \ar@{.}[r]_<{\txt{\tiny \{ Alice $\rightarrow$ \{ Coin $\rightarrow$ 5\}\}}} & \\
    \ar@{-->}[rr] ^{Time} & & & & & & & & & & &  &
}
\end{xy} }
\end{center}
\caption{Anomaly with naive definition of \emph{update-wins} map.}
\label{fig:map:updwins:error}
\end{figure*}
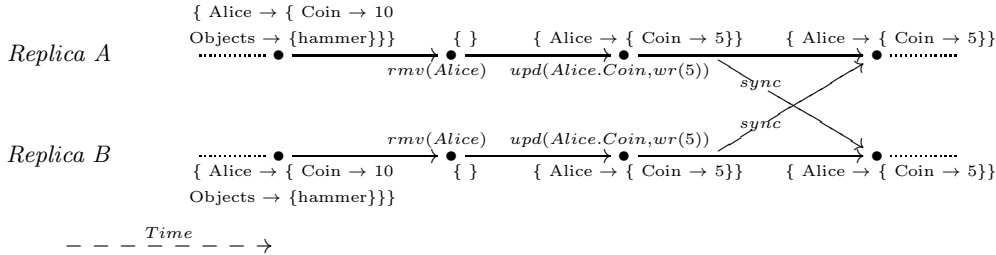

Precisely defining the concurrency semantics of an update-wins map is a bit more challenging.
Consider the example of Figure~\ref{fig:map:updwins:error}.
In this case, if the semantics was simply defined as an update canceling the effects of concurrent 
removes, the final value of the map would include the complete information for Alice, as the 
removes in both replicas would have no effects due to the concurrent updates.
 
%\begin{figure*}[h!]
%\begin{center}
%\footnotesize
%\begin{xy}
%%\xymatrixrowsep{6mm}
%%\xymatrixcolsep{6mm}
%\xymatrix{
%  \mathit{Replica\; A} & \ar@{.}[r] &
%  {\bullet} \ar@{->}[rr]^<{\txt{\tiny \{ sugar $\rightarrow$ 5\}}}_>{rmv(sugar)} & & 
%  {\bullet} \ar@{->}[rr]^<{\txt{\tiny \{ \}}}_>{\txt{\tiny upd(sugar,inc)}} & & 
%  {\bullet} \ar@{->}[rrr]^<{\txt{\tiny \{ sugar $\rightarrow$ 1\}}} & \ar@{->}[rrd] |(0.3){sync} & & 
%  {\bullet} \ar@{.}[r]^<{\txt{\tiny \{ sugar $\rightarrow$ 7 \}}} &\\
%  \mathit{Replica\; B} & \ar@{.}[r] &
%  {\bullet} \ar@{->}[rr]_<{\txt{\tiny \{ sugar $\rightarrow$ 5\}}}^>{\txt{\tiny rmv(sugar)}} & & 
%  {\bullet} \ar@{->}[rr]_<{\txt{\tiny \{ \}}}^>{\txt{\tiny upd(sugar,inc)}} & & 
%  {\bullet} \ar@{->}[rrr]_<{\txt{\tiny \{ sugar $\rightarrow$ 1\}}} & \ar@{->}[rru] |(0.3){sync} & & 
%  {\bullet} \ar@{.}[r]_<{\txt{\tiny \{ sugar $\rightarrow$ 7\}}} & \\
%%  \mathit{Replica\; C} & \ar@{.}[r] & \bullet \ar@{->}[rrr]^<{c_1}
%%& & &
%%  {\bullet} \ar@{->}[rr]^<{c_2}_<{\textit{Bored \ldots}} &
%%  & {\bullet} \ar@{.}[r]^<{c_3}_<{\textit{Can I join?}} 
%%& \\
%    \ar@{-->}[rr] ^{Time} & & & & & & & & &  
%}
%\end{xy} 
%\end{center}
%\caption{Anomaly with naive definition of \emph{update-wins} map.}
%\label{fig:map:updwins:error}
%\end{figure*}

We propose a different concurrency semantics for \emph{update-wins}.
Intuitively, the idea is that if at any moment, a key is removed in all replicas,
the updates on that key that happened before that state, in all replicas, will 
have no side-effects.
In our example, this means that the final state would have 5 Coins for Alice,
as this is the value written in the state of Alice, after she had been removed.

A little bit more formally, consider the transitive reduction of the 
happens-before graph of updates, which includes the edges that target 
a given update iff all source updates are concurrent among them.
In the reduced graph, for deciding which updates are relevant for computing the
state of key $k$, find the latest vertex-cut\footnote{We define the latest vertex-cut, $S$,
as one for which there is no other vertex-cut $S_i$ that includes an update that 
happened-after an update in $S$. Intuitively, this can be seen as the most recent vertex-cut,
considering the happens-before relation.} 
that includes only $\mathsf{rmv}(k)$ updates -- the relevant updates for computing the
value associated with $k$ are the ones that happened-after any of the 
$\mathsf{rmv}(k)$ updates in the vertex-cut. 
If there is no such cut, the effect of removes is canceled by concurrent updates, and
all updates (except removes) should be considered for determining the value of the map.

%==================================================================================================
%==================================================================================================
\subsubsection{Other CRDTs} 
A number of other CRDTs have been proposed in literature, including 
CRDTs for elementary data structures, such as Graphs \cite{Shapiro11Conflict}, and more
complex structures, such as JSON documents \cite{Kleppmann17JSON}.
For each of these CRDTs, the developers have defined and implemented a type specific 
concurrency semantics.

%==================================================================================================
%==================================================================================================
%==================================================================================================
\subsection{Discussion}

We now discuss several aspects related to the properties of concurrency semantics,
and how they related with the semantics under sequential execution.

\paragraph{Preservation of sequential semantics:}
When modeling an abstract data type that has an established semantics 
under sequential execution, CRDTs should preserve that semantics under a
sequential execution. 
For instance, CRDT sets should ensure that if the last update in a sequence 
of updates to a set added a given element, then a query immediately 
after that one will show the element to be present on the set. 
Conversely, if the last update removed an element, then a subsequent 
query should not show its presence. 

Sequential execution can occur even in distributed settings if 
synchronization is frequent. 
Replica A can be updated, merged into another replica B and updated there, 
and merged back into replica A before being updated again in replica A. 
In this case we have a sequential execution, even though updates 
have been executed in different replicas.

Historically, not all CRDT designs have met this property, typically by restricting
the functionality. 
For example, the \emph{two-phase set} CRDT \cite{Shapiro11Conflict} does not allow 
re-adding an element that was removed, and thus it breaks the common sequential 
semantics.

\paragraph{Principle of permutation equivalence:}
When defining the concurrency semantics for a given abstract data type, 
if all sequential permutations of updates
lead to the same state, then the final state of a CRDT under concurrent 
execution should also be that state
(principle of permutation equivalence \cite{Bieniusa12Semantics}).
As far as we know, all CRDTs proposed in literature that preserve the 
sequential semantics follow this principle\footnote{We note that it is possible
to design a CRDT that preserves sequential semantics but that does not
follow the principle of permutation equivalence.}.

\paragraph{Equivalence to a sequential execution:}
For some concurrency semantics, the state of a CRDT, as observed by executing 
a query, can be explained by a single sequential execution of the
updates (that might have been executed concurrently initially).
For example, the state of a LWW register (or LWW set) can be explained by
the sequential execution of all updates according to the total order 
defined among updates.

We note that this property differs from linearizable \cite{linearizability}, as
the property we are defining refers to a single replica and to the updates known 
at that replica, while linearizability concerns the complete system.

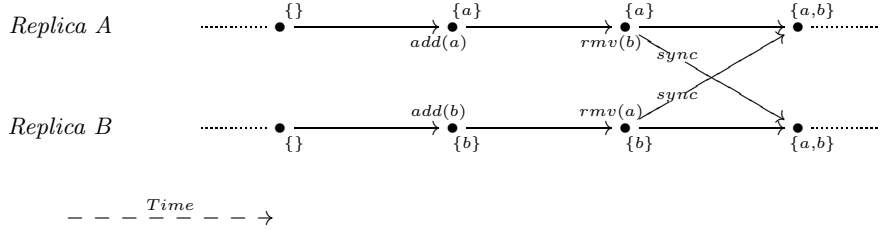
\begin{figure*}[h!]
\begin{center}
\footnotesize
\centerline{\begin{xy}
\xymatrix{
  \mathit{Replica\; A} & \ar@{.}[r] &
  {\bullet} \ar@{->}[rr]^<{\{\}}_>{add(a)} & & 
  {\bullet} \ar@{->}[rr]^<{\{a\}}_>{rmv(b)} & &
  {\bullet} \ar@{->}[rr]^<{\{a\}}\ar@{->}[rrd] |(0.3){sync} & & 
  {\bullet} \ar@{.}[r]^<{\{a,b\}} &\\
  \mathit{Replica\; B} & \ar@{.}[r] &
  {\bullet} \ar@{->}[rr]_<{\{\}}^>{add(b)} & &
  {\bullet} \ar@{->}[rr]_<{\{b\}}^>{rmv(a)} & &
  {\bullet} \ar@{->}[rr]_<{\{b\}} \ar@{->}[rru] |(0.3){sync} & & 
  {\bullet} \ar@{.}[r]_<{\{a,b\}} &\\
    \ar@{-->}[rr] ^{Time} & & &  & & & & &  
}
\end{xy} }
\end{center}
\caption{Add-wins set run showing that there might be no sequential execution of 
updates that explains CRDTs behavior.}
\label{fig:set:add-wins:noseq}
\end{figure*}

\paragraph{Extended behavior under concurrency:}
Not all CRDTs need or can be explained by sequential executions. 
The add-wins set is an example of a CRDT where there might be no sequential 
execution of updates that respect the happens-before relation to explain the state observed, 
as Figure~\ref{fig:set:add-wins:noseq} shows.
In this example, the state of the set after all updates propagate to all 
replicas includes $a$ and $b$, but in any sequential extension of the causal 
order a remove update would always be the last update, and consequently 
the removed element could not belong to the set.

Some other CRDTs can exhibit states that are only attained when concurrency 
does occur. An example is the \emph{multi-value register}, 
a register that supports
a simple write and read interface (see Section \ref{sec:appdev:conc_sem:register}). 
If used sequentially, sequential semantics is preserved, and a read will show the 
outcome of the most recent write in the sequence. 
However if two or more value are written concurrently, the subsequent read will show 
all those values (as the \emph{multi-value} name implies), and there is no sequential 
execution that can explain this result. 
We also note that a follow up write can overwrite both a single value and multiple values. 

\paragraph{Stability of arbitration:}
The concurrency semantics often includes arbitrating between concurrent 
updates, in which one update is given priority over some other, which 
will have no influence in determining the state of the object -- we call these
updates \emph{cast-off} updates. 
For example, in the add-wins set, an add update is given priority over concurrent 
remove updates.

A property that might influence both the implementation of the system and the 
way users reason about CRDTs is the stability of arbitration. 
Intuitively, we say that the arbitration is stable iff an update that was cast-off 
at some point, will remain cast-off in the future.

For defining this property more formally, we start by defining that two object 
states are observable equivalent if the result of executing any query in both states
is the same. 
An update $c$ is a cast-off update for a set of updates $O$ iff 
the states that result from applying the set of updates $O$ and 
$O \setminus \{c\}$ are observable equivalent\footnote{Note that this definition 
is more general than the intuitive definition, by considering as a cast-off update
one that has been made irrelevant by the execution of a subsequent update.}.
We say that the concurrency semantics defined for a CRDT is arbitration stable 
iff for any cast-off update $c$ in some set of updates $O$, 
when we consider a set of updates $O'$ that are concurrent or happened after all 
updates in $O$ ($\forall o' \in O' \mid \not \exists o \in O \cdot o' \prec o$), $c$ continues to 
be a cast-off update for the set of updates $O \cup O'$.

\begin{figure*}[h!]
\begin{center}
\footnotesize
\centerline{\begin{xy}
\xymatrix{
  \mathit{Replica\; A} & \ar@{.}[r] &
  {\bullet} \ar@{->}[rr]^<{0}_>{wr(4)} & & 
  {\bullet} \ar@{->}[rr]^<{4} & &
  {\bullet} \ar@{->}[rr]^<{5} \ar@{->}[rrd] |(0.3){sync} & & 
  {\bullet} \ar@{.}[r]^<{4} &\\
  \mathit{Replica\; B} & \ar@{.}[r] &
  {\bullet} \ar@{->}[rr]_<{0}^>{wr(5)} & &
  {\bullet} \ar@{->}[rr]_<{5}^>{wr(2)} \ar@{->}[rru] |(0.3){sync} & &
  {\bullet} \ar@{->}[rr]_<{2} \ar@{->}[rru] |(0.3){sync} & & 
  {\bullet} \ar@{.}[r]_<{4} &\\
    \ar@{-->}[rr] ^{Time} & & &  & & & & &  
}
\end{xy} }
\end{center}
\caption{Example of instability of arbitration in a register with data-driven conflict resolution.}
\label{fig:reg:order}
\end{figure*}
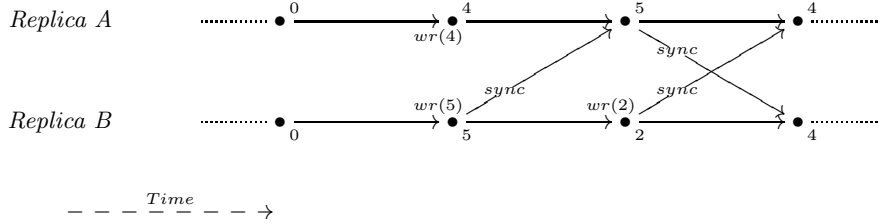

As an example, consider a register CRDT for which, in the presence of concurrent write updates,
the value of the register is that of the largest written value \cite{Zawirski:2016:ECR:2911151.2911157}.
More precisely, for a set of updates $O$, let $O_{max}$ be the causal frontier 
defined as the set of updates 
for which there is no subsequent update in the happens-before order, i.e., 
$O_{max} =\{o \in O \mid \not \exists o' \in O \cdot o \prec o'\}$.
The value of the set is $max(\{ v \mid wr(v) \in O_{max}\})$.
In the run presented in Figure \ref{fig:reg:order}, replica A, after receiving
the update $\mathsf{wr}(5)$ from replica B arbitrates that $\mathsf{wr}(5)$ should
be prioritized over $\mathsf{wr}(4)$ because the written value is larger (and 
$\mathsf{wr}(4)$ becomes a cast-off update). 
However, after later receiving $\mathsf{wr}(2)$, $\mathsf{wr}(5)$ no longer belong 
to the $O_{max}$ set of $O$ and thus $\mathsf{wr}(4)$ should be given priority.

The instability of arbitration raises two issues.
First, the state in replica A evolves in a somehow unexpected way for an observer, 
as the integration of a remote update makes the object move to a prior value.
Second, for an implementation it may have implications on when the information about
an update can be discarded -- the information of a cast-off update cannot be discarded
if the update may later become relevant (not cast-off) \footnote{The original design
of the register with data-driven conflict resolution \cite{Zawirski:2016:ECR:2911151.2911157} 
was incorrect because it discarded cast-off operations before it is safe doing it.}.

Although we used a simple register for exemplifying the instability of arbitration,
we conjecture that this issue will occur for any CRDT in which the value is 
decided by arbitrating over the updates in the causal frontier of updates and there
are more than two possible values. 
We further conjecture that in CRDTs that exhibit arbitration instability, 
a cast-off update cannot be discarded in a replica before the replica has received all
concurrent updates. 

\subsection{Advanced topics}

In this section we discuss two topics that are relevant for the application
programmer and that have not been addressed extensively in literature.

\subsubsection{Concurrency semantics with multiple options} 

The concurrency semantics defines the behavior of the CRDT in the
presence of concurrent updates. 
For example, considering the concurrency semantics presented in Section \ref{sec:appdev:conc_sem},
when an application developer wants to use a set, she must select between 
\emph{add-wins}, \emph{remove-wins} and \emph{last-writer-wins} set.

However, sometimes having a single semantics is not enough, as in this example
from Rijo et. al. \cite{addremovewins}:

\begin{quotation}
Consider a set that maintains the users who are currently in a chat room. 
A \emph{remove} is executed by a replica that detects that the connection to the user is lost. 
An \emph{add} is executed by a replica that detects a (re)connection from a user. 
With an \emph{add-wins} policy, when a user migrates from one replica 
to the other and concurrent add and remove are executed, the user will remain in the set.
However, when the user does a \emph{logout}, it makes sense that the 
user is removed from the set even if concurrent \emph{adds} happen. 
\end{quotation}

This example shows that, sometimes, it would be interesting to have 
a CRDT where the concurrency semantics would allow different final results
depending on the situation. 
A similar point, on the interest of having multiple policies for handling concurrent 
updates in the same CRDT, was made by Ivanov et. al. \cite{ivanov}.
Rijo et. al. \cite{addremovewins} addressed this challenge, by proposing 
a set CRDT that exports two alternative remove operations, a default 
remove and a strong remove.
The concurrency semantics states that in the presence of concurrent updates
for the same element, an add update wins over concurrent default removes and 
a strong remove wins over concurrent adds. 

This allows to address the problem presented\footnote{Note that although the 
proposed design exhibits arbitration instability, this is not an issue in this
case because a replica that has seen that the user has logged-out will 
not automatically add her again.}.
However, addressing this issue in a more general way is an open research problem.

\subsubsection{Encapsulating invariant preservation} 

CRDTs encapsulate the merge of concurrent updates, making complex concurrency
semantics available to any application programmer.
As mentioned in the introduction, some applications require global invariants
to be maintained for correctness.
The CRDTs discussed in Section~\ref{sec:appdev:conc_sem} are unable to enforce
such global invariants.
We now discuss how some global invariants can be enforced by encapsulating the
necessary algorithms inside a CRDT, 
thus empowering application developers with much more powerful abstractions.

It is known that some global invariants can be enforced under weak 
consistency \cite{Bailis14Coordination}.
Other invariants, such as numeric invariants that usually are enforced with 
global coordination, can be enforced in a conflict free manner by using 
escrow techniques \cite{ONeil86Escrow} that split the available resources
by the different replicas.  
The Bounded Counter CRDT \cite{DBLP:conf/srds/BalegasSDFSRP15} defines a counter that never goes 
negative, by encapsulating an implementation of escrow techniques that
runs under the same system model of CRDTs. 
The Bounded Counter assigns to each replica a number of allowed decrements under the condition that 
the sum of all allowed decrements do not exceed the value of the counter.
While its assigned decrements are not exhausted, replicas can accept decrements without coordinating
with other replicas.
After a replica exhaust its allowed decrements, a new decrement will either fail or require synchronizing 
with some replica that still can decrement.

It is possible to generalize this approach to enforce other system wide 
invariants, including invariants that enforce conditions over the state 
of multiple CRDTs \cite{Balegas15Putting}. 
An interesting research question is which other invariants can be enforced
(e.g. by adapting other algorithms proposed in the past \cite{Barbara94Demarcation,Walborn95Supporting})
and what other functionality can be encapsulated in objects that are updated
using an eventual consistency model.

\subsubsection{Transactions and other programming models}\label{sec:appdev:adv:tx}
CRDTs have been used in several storage systems that provide different forms of 
transactions.

SwiftCloud \cite{swiftcloud} and Antidote \cite{Akkoorath16Cure}
provide a weak form of transactions that is highly available \cite{hat} and 
never abort, with reads observing a snapshot of the database and concurrent writes 
being merged using CRDT's defined concurrency semantics.   

Walter \cite{Sovran11Transactional} and Sieve \cite{Li14Automating} provide
support for both weak and strong forms of transactions. 
While weak transactions can execute concurrently, with concurrent updates 
being merged using CRDTs, strong transactions (that may access non-CRDT objects)
are executed according to a serial order.

Lasp \cite{lasp} proposes a dataflow programming-model where the state of a CRDT
is the result of executing a computation over the state of some other CRDTs. 
Those former CRDTs are similar to views in database systems, and the Lasp runtime 
executes an algorithm that tries to achieve a goal similar to that of incremental 
materialized view maintenance in database systems \cite{Gupta99Maintenance}.

\section{CRDTs for the System Developer}\label{sec:sysdev}

A system developer using CRDTs needs to create a distributed system that integrates 
some CRDT implementations. Unlike the application developer, that is mostly concerned with the
functionality provided by the CRDT, the system developer focus is on how to guarantee
that the replicas of CRDTs in the multiple nodes of the system are kept synchronized.
As CRDTs guarantee that all replicas converge to the same state when all updates
propagate to all replicas, the system developer needs to focus on guaranteeing that
all updates reach all replicas -- we call this aspect, the synchronization model.

\subsection{Synchronization model}

\subsubsection{State-based synchronization}\label{sec:sysdev:state}
In state-based synchronization, replicas synchronize by establishing bi-direc\-tional (or
unidirectional) synchronization sessions, where both (one, resp.) replicas send
their state to a peer replica. 
When a replica receives the state of a peer, it merges the received state with 
its local state.

CRDTs designed for state-based replication define a merge function to 
integrate the state of a remote replica.
It has been shown \cite{Shapiro11Conflict} that all replicas of a CRDT converge if: 
\begin{inparaenum}[(i)] 
\item the possible states of the CRDT are partially ordered according to $\leq$ forming a 
join semilattice;
\item an update modifies the state $s$ of a replica by an inflation, producing a new state that is
larger or equal to the original state according to $\leq$, 
%i.e., for any operation $u$, $s \leq s \bullet u$;
  i.e., for any update $u$, $s \leq u(s)$;
\item the merge function produces the join (least upper bound) of two states, 
i.e. for states $s_1,s_2$ it derives $s_1 \sqcup s_2$. 
\end{inparaenum}

For guaranteeing that all updates eventually reach all replicas, it is only necessary 
to guarantee that the synchronization graph is connected.
A number of replicated systems \cite{lotusnotes,Ratner99Peer,dynamo} used this approach,
adopting different synchronization schedules, topologies and mechanisms 
for deciding if and what data to exchange.

\subsubsection{Operation-based synchronization}

In operation-based synchronization, replicas converge by propagating updates 
to every other replica.
When an update is received in a replica, it is applied to the local replica
state.
Besides requiring that every update operation is reliably delivered to all replicas,
some CRDT designs require updates to be delivered according to some specific order,
with causal order being the most common.

CRDTs designed for operation-based replication must define, for each update, 
a generator and an effector function. 
The generator function executes in the replica 
where the update is submitted, it has no side-effects and generates an effector 
that encodes the side-effects of the update. In other words, the effector is a 
closure created by the generator depending on the state of the origin replica.
The effector operation must be reliably executed in all replicas, where it
updates the replica state. 
It has been shown \cite{Shapiro11Conflict} that if effector operations are delivered
in causal order, replicas will converge to the same state if concurrent effector operations
commute.
If effector operations may be delivered without respecting causal order, then all 
effector operations must commute.
If effector operation may be delivered more than once, then all effector operations
must be idempotent.
Most operation-based CRDT designs require exactly-once and causal delivery. 

To guarantee that all updates are reliably delivered to 
all replicas, it is possible to rely on any reliable multicast communication 
subsystem.
A large number of protocols have been proposed for achieving this 
goal \cite{Chang84Reliable,Birman87Reliable,Demers87Epidemic}. 
In practice, it is also possible to rely on a reliable publish-subscribe 
system, such as Apache Kafka\footnote{https://kafka.apache.org/} for this purpose.

\subsubsection{Alternative synchronization models}

\paragraph{Delta-based synchronization:}
When comparing state-based and opera\-tion-based synchronization, a simple
observation is that if an update only modifies part of the state, 
propagating the complete state for synchronization to a remote replica is 
inefficient, as the remote replica already knows most of the state. 
On the other hand, if a large number of updates modify the same 
state, e.g. increments in a counter, propagating the state once is much more
efficient than propagating all update operations.
To address this issue, it is possible to design CRDTs that allow being 
synchronized using both state-based and operation-based approaches \cite{Bieniusa12Optimized}. 

Delta-state CRDTs \cite{DBLP:conf/netys/AlmeidaSB15,DBLP:journals/jpdc/AlmeidaSB18} address this issue 
in a principled way by propagating delta-mutators, that encode the changes that have been made
to a replica since the last communication.
The first time a replica communicates with some other replica, the full state
needs to be propagated.
Big delta state CRDTs \cite{vanderLinde06bigdelta} improve the cost of 
first synchronization by being able to compute delta-mutators from a summary
of the remote state. 
Join-decompositions \cite{Enes2017} are an alternative approach to achieve the same goal 
by computing digests that help determining which parts of a remote state are needed.

In the context of operation-based replication, effector operations should be applied 
immediately in the source replica, that executed the generator. 
However, propagation to other replicas can be deferred for some period and 
effectors stored in an outbound log, presenting an opportunity to 
compress the log by rewriting some operations -- e.g. two $\mathsf{add}(1)$ operations
in a counter can be converted in a $\mathsf{add}(2)$ operation. 
Several systems \cite{rover,Kistler92Disconnected,Preguica00Data,Cabrita17Non} have included
mechanisms for compressing the log of operations used for replication.
We note that delta-mutators can also be seen as a compressed representation
of a log of operations.

\paragraph{Pure operation-based synchronization:}
The operation-based CRDTs require executing a generator function against
the replica state to compute an effector operation.
In some scenarios, this may introduce an unacceptable delay for propagating
an update. 
Pure-operation based CRDTs \cite{Baquero14Making} address this issue by 
allowing the original update operations to be propagated to all replicas,
typically at the cost of more complex operation representation and 
of having to store more metadata in the CRDT state.

\subsubsection{Discussion}

In this section we identify a number of situations and discuss how to address
them in each of the synchronization models.

\paragraph{Immediate synchronization required:} In systems where updates
must be propagated immediately to other replicas, state-based synchronization 
is the worst choice, as it is more costly to propagate the state than a
single operation. 
Delta-based synchronization is still a choice, but in this
case propagating delta-mutators is equivalent to propagate operations (as
there is no chance of compressing multiple operations into a single delta-mutator) 
with the difference that delta-mutators are idempotent and do not require 
exactly-once delivery (see discussion next).

\paragraph{Replica synchronization after failures:} State-based and delta-based 
synchronization allow a replica to synchronize with some other replica by receiving its
state. In this case, both state-based and delta-based are equivalent.
Improvements proposed to the delta-based approach can help minimizing the 
state that needs to be transferred.

For solutions based on operations, possible solutions include:
\begin{inparaenum}[(i)]
\item using the local state, and replaying the missing operations; or
\item copying the state from a remote replica, and replaying the missing operations.
\end{inparaenum}

\paragraph{Idempotence and exactly-once delivery:} Operation-based CRDTs 
typically require an operation to be delivered exactly-once to every replica.
If operations are idempotent, it is possible to drop this requirement, but
designing idempotent operations is often more complex. 
In practice, enforcing exactly-once delivery is usually straightforward, 
as the metadata used to guarantee reliable delivery of operations (a summary
of operations received) can also be used to detect if an operation has 
already been delivered at a replica or not. 
Thus, although dropping the exactly-once delivery requirement might be an
interesting conceptual property, it might be less relevant in practice and
not worth if it leads to more complex operation designs.

\subsection{Applications}

CRDTs have been used in a large number of distributed systems and applications
that adopt weak consistency models. 
The adoption of CRDTs simplifies the development of these systems and applications, 
as CRDTs guarantee that replicas converge to the same state when all updates are 
propagated to all replicas.
We can group the systems and applications that use CRDTs into two groups: storage
systems that provide CRDTs as their data model; and applications that embed CRDTs
to maintain their internal data.

CRDTs have been integrated in several storage systems that make them available 
to applications.
An application uses these CRDTs to store their data, being the responsibility of 
the storage systems to synchronize the multiple replicas.
The following commercial systems use CRDTs: Riak \footnote{Developing with Riak KV 
Data Types \url{http://docs.basho.com/riak/kv/2.2.3/developing/data-types/}.}, Redis \cite{rediscrdts} and Akka \footnote{Akka Distributed Data: \url{https://doc.akka.io/docs/akka/2.5.4/scala/distributed-data.html}.}.
A number of research prototypes have also used CRDTs, including Walter~\cite{Sovran11Transactional},
SwiftCloud~\cite{swiftcloud} and AntidoteDB\footnote{AntidoteDB: \url{http://antidotedb.org/}.}~\cite{Akkoorath16Cure}.

CRDTs have also been embedded in multiple applications. 
In this case, developers either used one of the available CRDT libraries,
implemented themselves some previously proposed design or designed new
CRDTs to meet their specific requirements. 
An example of this latter use is Roshi\footnote{Roshi is a large-scale CRDT set implementation for timestamped events: \url{https://github.com/soundcloud/roshi}.}, a LWW-element-set CRDT used 
for maintaining an index in SoundCloud stream.

\subsection{Advanced topics}

In this section we discuss other topics that may be important for a 
system developer. Some topics that are relevant for both the system developer
and the CRDT developer will be addressed in the next section.

\subsubsection{CRDTs and transactions}\label{sec:sysdev:adv:tx}

As mentioned in Section \ref{sec:appdev:adv:tx}, some database systems
include support for weak forms of transactions that execute queries and
updates in CRDTs.
A number of transactional protocols have been proposed in literature
\cite{swiftcloud,Akkoorath16Cure,Sovran11Transactional}.

The use of CRDTs in these transactional protocols raises two main challenges.
First, all updates executed by concurrent transactions need to be 
considered to define the final state of each CRDT. 
The protocols implemented in most systems 
\cite{swiftcloud,Akkoorath16Cure,Sovran11Transactional,Li14Automating} use operation-based CRDTs,
with all updates being replayed in all replicas.
Second, in some geo-replicated systems \cite{swiftcloud,Akkoorath16Cure}, 
it is possible that new updates are received from remote replicas while a local transaction is running.
To allow transactions to access a consistent snapshot, the system must
be able to maintain multiple versions for each object.

Some of these database systems \cite{Sovran11Transactional,Li14Automating} 
also provide support for strong forms of
transaction by executing protocols that guarantee that these transaction execute
in a serial order.
Droppel \cite{droppel} provide serializability by combining a multi-phase 
algorithm where some operations are only allowed in one of the phase, 
with constructs similar to CRDTs for allowing 
the concurrent access of multiple transactions to the same objects.

\subsubsection{Support for large objects}

Creating CRDTs that can be used for storing complex application data 
may simplify application development, but it can lead to performance
problems as the system needs to handle these potentially  
large data objects.
This problem occurs both in the servers, as a small update to a 
large object may lead to loading and storing large amounts of data
from disk, and when transferring CRDTs to clients, as large objects
may be transferred when only a part of the data is necessary.

To address this problem, for objects that are compositions of
other objects, instead of having a single object that embeds all other
objects, it is possible to maintain each object separately and use references 
to link the object~\cite{Meiklejohn14Composability} -- e.g. in a map, an object 
associated with a key would be stored separately and the map would have 
only a reference to it.
To guarantee the correctness of replication for application updates that 
span multiple objects, the system may need to guarantee that updates are
executed according to some specific ordering or atomically, depending 
on the operations. 
For example, when associating a new object with a key, it is necessary that
the new object is created in all replicas before adding the reference to 
it. This can be achieved by resorting to causal 
consistency \cite{cops,chainreaction,swiftcloud} or to atomic updates (as discussed 
in Section~\ref{sec:sysdev:adv:tx}).

Sometimes objects are large because they hold large amounts of information -- e.g. 
a set with a very large number of elements. 
In this case, the same performance problems may arise. 
A possible solution to address this problem is to split the object
in multiple sub-objects. For example, a set can be implemented by maintaining
multiple subsets with disjoint information and a simple root object that only keeps
references to these subsets. When adding, removing or querying for an element, the 
operation should be forwarded to the subset that holds the given element.  
This approach has been adopted in the Riak database for efficiently supporting
sets and maps with large amounts of information. 
Briquemont et. al. \cite{Briquemont15Conflict} have proposed a principled approach 
to define partial replicas adopting this idea.

\subsubsection{Non-uniform replicas}
The replication of CRDTs typically assumes that eventually all replicas
will reach the same state, storing exactly the same data.
However, depending on the read operations available in the CRDT interface, 
it might not be necessary to maintain the same state in all replicas.
For example, an object that has a single read operation returning the 
top-K elements added to the object only needs to maintain those top-K elements 
in every replica. 
The remaining elements are necessary if a remove operation is available,
as one of the elements not in the top needs to be promoted when a top element 
is removed. 
Thus, each replica can maintain only the top-K elements and the elements
added locally.

This replication model is named non-uniform replication \cite{Cabrita17Non}
and can be used to design CRDTs that exhibit important storage and bandwidth 
savings when compared with alternatives that keep all data in all replicas.
Although it is clear that this model cannot be used for all data types, 
several useful CRDT designs have been proposed, including top-K, top-Sum and
histogram. To understand what data types can adopt this model and how to 
explore it in practice is an open research question.
Non-uniform replication is both an alternative and a complementary approach 
to the solutions that maintain partial replicas.

\subsubsection{CRDTs and storage}
For systems that need to save operation-based CRDTs on stable storage,
it might be interesting to store a stable version and a sequence of updates
until the CRDT is read. 
This allows to store a new update without having to read the current state of
the object, minimizing the overhead imposed by maintaining objects
that are updated by executing operations when compared with simple registers
where an update operations simply overwrites the previous value.

Some storage systems, such as RocksDB\footnote{RocksDB: http://rocksdb.org/} and
FASTER, provide native support for such functionality (often called  
read-modify-write updates). For example, in RocksDB \footnote{RocksDB: http://rocksdb.org/},
the programmer can specify a merge operator to be used by the system when:
\begin{inparaenum}[(i)]
\item it needs to create the current version, by applying a sequence of updates
to the latest materialized version; and
\item it needs to compress multiple operations into a single one.
\end{inparaenum}
The base system includes a Counter that is similar to an operation-based counter 
CRDT.
\note{We have implemented a database that stores other CRDTs on top of RocksDB
by exploring this functionality: \url{https://github.com/preguica/RocksDBCRDTDB}.}

\section{CRDTs for the CRDT Developer}\label{sec:crdtdev}

This section addresses CRDTs from the point of view of a CRDT developer.
We start by discussing techniques for designing CRDTs that implement
the concurrency semantics discussed in Section \ref{sec:appdev:conc_sem}.

\subsection{Techniques for designing CRDTs}

As shown by Burckhardt et. al. \cite{Burckhardt14Replicated}, a CRDT can be
specified by relying on:
\begin{inparaenum}[(i)]
\item the full history of updates executed; 
\item the happens-before relation among updates; and 
\item an arbitration relation among updates (when necessary).
\end{inparaenum}
A query can be specified as a function that uses this information and
the value of parameters to compute the result.
Given this, it is possible to implement a CRDT by just storing 
this information and synchronize replicas either using a state-based or 
operation-based approach. 

However, this implementation would be rather inefficient. 
The goal of an actual implementation is to minimize the data that needs 
to be stored in each replica and the data that needs to be transferred for
synchronizing replicas.

The goal of this section is not to provide an exhaustive catalog of CRDT 
implementations, but only to introduce and explain the main techniques that 
have been used to design CRDTs, giving examples of their use.

\paragraph{Commutativity and associativity:}
For some CRDTs, the updates defined are intrinsically commutative and associative.
In this case, defining an operation-based CRDT is straightforward, as each replica
only need to maintain the value of the CRDT, and updates simply modify the value of the replica.
Algorithm \ref{alg:op:counter} exemplifies the case with an operation-based counter.
The state is just an integer with the current value of the counter. 
The increment update receives as parameter the value to increment (that can be negative)
and adds it to the value of the counter in the effector operation.

\begin{algorithm}
\caption{Operation-based Counter CRDT (adapted from \cite{Shapiro11Comprehensive}).}
\label{alg:op:counter}
\begin{algorithmic}[1]
\Payload{int $\mathit{val}$}\Comment{Initial value: 0}
\EndPayload
\Query{value}{}{int}
\State \Return val
\EndQuery
\UpdateOp{inc}
\AtSource{int \emph{n}}
\State \textbf{return} $(\mathsf{inc}, [n])$ \Comment{Operation name and parameters for effector}
\EndAtSource
\Downstream{int \emph{n}}
\State $\mathit{val}$ := $\mathit{val}$ + $n$
\EndDownstream
\EndUpdateOp
\end{algorithmic}
\end{algorithm}

A state-based implementation can rely on the commutativity and associativity properties
for computing for each replica a partial value, and aggregate the partial
values into the global value. 
Algorithm \ref{alg:st:counter} presents a state-based counter CRDT.
The state of the CRDT consists of two associative arrays, 
one with the sum of positive values added and the other with the sum of
negative values added in each replica. 
For each replica, these arrays maintain a partial result that reflects the 
updates submitted in that replica -- this is only possible because the function being
computed is associative.

Unlike the operation-based implementation, it is necessary to keep the positive 
and negative values added in separate arrays to guarantee that the Counter respects the
properties for being a state-based CRDT.
This is necessary because when merging it is necessary to keep, for each entry
in the array, the most recent value. 
By keeping two arrays, it is known that the most recent value for a given entry is
the one with the largest absolute value, as the absolute value for an entry never 
decreases.
This would not be the case if a single array was used, as the value of an entry would
increase or decrease as the result of executing an $\textsf{inc}$ with a positive and 
value.

\begin{algorithm}
\caption{State-based Counter CRDT (adapted from \cite{Shapiro11Comprehensive}).}
\label{alg:st:counter}
\begin{algorithmic}[1]
\Payload{int[] $\mathit{valP}$, int[] $\mathit{valN}$} \Comment{For any id, the initial value is $0$}
\EndPayload
\Query{value}{}{int}
\State \Return $\sum_r \mathit{valP}[r]$ + $\sum_i \mathit{valN}[r]$
\EndQuery
\Update{inc}{int \emph{n}}
    \State \Let $\mathit{id}$ := \emph{repId}() \Comment{\emph{repId}: generates the local replica id}
	\If{$n > 0$}
    \State $\mathit{valP}[\mathit{id}]$ := $\mathit{valP}[\mathit{id}] + n$
    \Else
    \State $\mathit{valN}[\mathit{id}]$ := $\mathit{valN}[\mathit{id}] + n$
    \EndIf
\EndUpdate
\Merge{X}{Y}{Z}
	\For{$r \in X.\mathit{valP}.\mathit{keys} \cup X.\mathit{valN}.\mathit{keys} \cup Y.\mathit{valP}.\mathit{keys} \cup Y.\mathit{valN}.\mathit{keys}$}
    \State {$Z.\mathit{valP}[r]$ := $max(X.\mathit{valP}[r], Y.\mathit{valP}[r])$}
    \State {$Z.\mathit{valN}[r]$ := $min(X.\mathit{valN}[r], Y.\mathit{valN}[r])$}
    \EndFor
\EndMerge
\end{algorithmic}
\end{algorithm}

%\begin{algorithm}
%\caption{State-based Counter CRDT}
%\label{alg:op:counter}
%\begin{algorithmic}[1]
%\Payload{int[] $val$, int[] $vrs$}
%\EndPayload
%\Query{value}{}{int}
%\State \Return $\sum_i val[i]$
%\EndQuery
%\Update{inc}{int \emph{n}}
%    \State \Let $id$ := \emph{repId}() \Comment{\emph{repId}: generates the replica id}
%    \State $val[id]$ := $val[id] + n$
%    \State $vrs[id]$ := $vrs[id] + 1$
%\EndUpdate
%\Merge{X}{Y}{Z}
%	\For{$k \in X.vrs.keys \cup Y.vrs.keys$}
%	\If{$X.vrs[k] > Y.vrs[k]$}
%    \State \Let {$Z.val[k]$ := $X.val[k])$}
%    \Else
%    \State \Let {$Z.val[k]$ := $Y.val[k])$}
%	\EndIf
%    \State \Let {$Z.vrs[k]$ := $max(X.vrs[k],Y.vrs[k])$}
%    \EndFor
%\EndMerge
%\end{algorithmic}
%\end{algorithm}

\paragraph{Total order for arbitration:}
As discussed in Section \ref{sec:appdev:conc_sem}, some CRDTs use a
total order among updates to arbitrate the value of the CRDT.
In this case, the CRDT needs to record the information necessary to do
the arbitration.

Algorithm \ref{alg:st:lwwregister} presents a state-based LWW register CRDT. 
The state of the CRDT includes a value and a timestamp, with the
timestamp being used for arbitration.
On a write update, besides updating $val$, the timestamp $ts$ is
assigned the current timestamp -- we assume the timestamps generated are 
totally ordered and compatible with the happens-before relation (as discussed in Section \ref{sec:appdev:conc_sem}).
When merging two replicas, the timestamps are compared to decide
which value to keep -- the one of the replica with the largest timestamp.

\begin{algorithm}
\caption{State-based LWW Register CRDT (adapted from \cite{Shapiro11Comprehensive}).}
\label{alg:st:lwwregister}
\begin{algorithmic}[1]
\Payload{T $\mathit{val}$, timestamp $ts$} \Comment{Initial value: $\bot,\bot$}
\EndPayload
\Query{value}{}{T}
\State \Return $\mathit{val}$
\EndQuery
\Update{wr}{T \emph{v}}
    \State $\mathit{ts}$ := \emph{curTimestamp}() \Comment{Totally ordered timestamp}
    \State $\mathit{val}$ := $v$
\EndUpdate
\Merge{X}{Y}{Z}
	\If{$X.ts > Y.ts$}
    \State $Z.\mathit{val}, Z.\mathit{ts}$ := $X.\mathit{val}, X.\mathit{ts}$
    \Else
    \State $Z.\mathit{val}, Z.\mathit{ts}$ := $Y.\mathit{val}, Y.\mathit{ts}$
    \EndIf
\EndMerge
\end{algorithmic}
\end{algorithm}

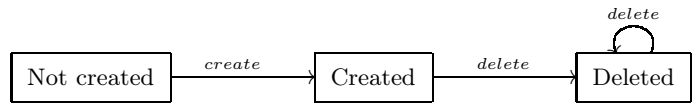
\begin{figure*}[h!]
\begin{center}
\footnotesize
\centerline{\begin{xy}
\xymatrix{
	*++[F]{\txt{Not created}} \ar@{->}[rr]^{create} && 
	*++[F]{\txt{Created}} \ar@{->}[rr]^{delete} && 
	*++[F]{\txt{Deleted}} \ar@{->}@(ur,ul)_{delete}
}
\end{xy} }
\end{center}
\caption{Life-cycle of a unique identifier.}
\label{fig:uids}
\end{figure*}

\paragraph{Unique identifiers:}
Many CRDT designs rely heavily on the use of unique identifiers.
In this context, the use of unique identifiers follows a very simple life-cycle, 
presented in Figure \ref{fig:uids}, that exhibits several interesting properties.
First, the creation of a unique identifier is done without coordination and
it is impossible to create the same unique identifier more than once -
a single create moves a unique identifier from not being created to being created.
Second, the creation of the unique identifier always happens before any action 
that refers the unique identifier. 
Third, after a unique identifier is created and used in a CRDT, the only operation 
that can be performed is to delete it. Multiple deletes can be executed, but all
deletes lead to the same deleted state.
For any given unique identifier, a create always happens-before any delete.

Algorithm \ref{alg:op:addwins} present the code of an operation-based add-wins set CRDT.
This is a good example of how unique identifiers are used -- the key information for the 
design are the unique identifiers.
Whenever a new element is added, a unique identifier is created and associated with the 
added element.
A remove removes only elements for which the add happened before the remove -- basically
this consists in deleting the unique identifiers associated with the element 
that are known at the moment of the remove
(this is the reason why the add-wins set was originally called observed-remove set).
Thus, unique identifiers created concurrently with a remove are not deleted, leading 
to the add-wins semantics.

\begin{algorithm}
\caption{Operation-based Add-wins Set CRDT (adapted from \cite{Shapiro11Comprehensive})}
\label{alg:op:addwins}
\begin{algorithmic}[1]
\Payload{set $S$} \Comment{Set of pairs (element, unique-id); initial value: $\emptyset$}
\EndPayload
\Query{lookup}{element e}{boolean}
\State \Return ($\exists (e,u) \in S$)
\EndQuery
\UpdateOp{add}
\AtSource{element \emph{e}}
\State \textbf{return} $(\mathsf{add}, [e,\mathit{newUID}()])$
\EndAtSource
\Downstream{element \emph{e}, unique-id \emph{u}}
\State $S$ := $S \cup \{(e,u)\}$
\EndDownstream
\EndUpdateOp
\UpdateOp{rmv}
\AtSource{element \emph{e}}
\State \textbf{return} $(\mathsf{rmv}, [e,\{u \mid (e,u) \in S\}])$
\EndAtSource
\Downstream{element \emph{e}, set uids}
\State $S$ := $S \setminus \{(e,u) \mid u \in uids\}$
\EndDownstream
\EndUpdateOp
\end{algorithmic}
\end{algorithm}

\paragraph{Summaries of unique identifiers:}
For the design of state-based CRDTs, the life-cycle of unique identifiers has a problem:
it is impossible to distinguish between the not created and deleted states.
Thus, it is not clear what to do when merging two replicas that differ only by 
the fact that one includes information associated with a unique identifier 
and the other that does not include such information.
If the unique identifier was already deleted, the latter is the most recent
version; if not, the most recent version would be the former.
To address this issue, some initial state-based CRDT designs recorded deleted unique 
identifiers as tombstones.

A more ingineous approach was proposed by Bieniusa et. al. \cite{Bieniusa12Optimized}.
The key idea is to have an efficient summary of unique identifiers observed
in each replica. With this summary, when a replica does not maintain information
for a unique identifier but that identifier is in the summary of known identifier,
then it is because the information for the unique identifiers has been deleted.
This leaves the question of how to create unique identifiers that can be 
efficiently summarized. 
An easy solution for this is to use a pair (timestamp, replica identifier)
and summarize these pairs in a vector clock\footnote{We note that if the 
timestamp is a Lamport Clock \cite{Lamport78Time} (or an 
hybrid clock), it is even possible to establish a total order compatible with 
the happens-before relation among identifiers by defining that $(t_1,r_1) < (t_2,r_2)$
iff $t_1 < t_2 \vee ( t_1 = t_2 \wedge r_1 < r_2)$.}.

Figure \ref{alg:st:addwins} presents the design of a state-based add-wins
set that adopts this approach. 
The summary of unique identifiers is stored in the vector clock $vv$, where
each entry has the largest timestamp known for the given replica (a value of 
$n_r$ for replica $r$ means that unique identifiers $(1,r),\ldots,(n_r,r)$ have been
observed).
In a given replica, the new unique identifier is created by incrementing the last
generated timestamp, and using the new value in combination with the replica identifier.
The add operation generates a new unique identifier and records a pair that
associates it with the added element.
The remove operation removes all pairs for the removed element.
The merge function computes the merged state by taking the union of the pairs
in each of the replicas that have not been deleted in the other replica --- these
pairs are those that are presented in replica $r_1$ (resp. $r_3$) and not in 
replica $r_2$ (resp. $r_1$), but for which the
unique identifier is reflected in $vv$ of $r_2$ (resp. $r_1$).

\begin{algorithm}
\caption{State-based Add-wins Set CRDT (adapted from \cite{Bieniusa12Optimized})}
\label{alg:st:addwins}
\begin{algorithmic}[1]
\Payload{set $S$, int[] $\mathit{vv}$} \\
\Comment{S: set of pairs (element e, (timestamp t, rep\_id r)); initial value: $\emptyset$}\\
\Comment{vv: summary of observed unique ids; for any id, the initial value is 0}
\EndPayload
\Query{lookup}{element e}{boolean}
\State \Return ($\exists (e,u) \in S$)
\EndQuery
\Update{add}{element \emph{e}}
\State \Let $r$ = $\mathit{getReplicaID}()$
\State $\mathit{vv}[r]$ := $\mathit{vv}[r] + 1$ 
\State $S$ := $S \cup \{(e,(\mathit{vv}[r],r))\}$ 
\EndUpdate
\Update{rmv}{element \emph{e}}
\State $S$ = $S \setminus \{(e,(t,r)) \in S\}$
\EndUpdate
\Merge{X}{Y}{Z}\\
\Comment{Elements that are in X, but that have been deleted in Y}
\State \Let $\mathit{RX}$ := $\{(e,(t,r)) \in X.S \mid (e,(t,r)) \not \in Y.S \wedge Y.\mathit{vv}[r] \geq t\}$\\
\Comment{Elements that are in Y, but that have been deleted in X}
\State \Let $\mathit{RY}$ := $\{(e,(t,r)) \in Y.S \mid (e,(t,r)) \not \in X.S \wedge X.\mathit{vv}[r] \geq t\}$
\State $Z.S$ = $(X.S \setminus \mathit{RX}) \cup (Y.S \setminus \mathit{RY})$ 
\State $Z.\mathit{vv}$ = $\mathit{max}_{\mathit{pointwise}}(X.\mathit{vv},Y.\mathit{vv})$ 
\EndMerge
\end{algorithmic}
\end{algorithm}

\paragraph{Dense space for unique identifiers:}
In the list CRDT discussed in Section \ref{sec:appdev:conc_sem}, an element is inserted
between two other elements. 
Some list CRDT designs \cite{Preguica09Commutative,Weiss09Logoot} rely on using 
unique identifiers that encode the relative position among elements. Thus, for 
adding a new element between elements $prv$ and $nxt$, it is necessary to generate 
a unique identifier that will be ordered between the unique identifiers of the 
elements $prv$ and $nxt$. 
This requires being able to always create a unique identifier between two existing 
unique identifiers, i.e., to have a dense space of unique identifiers.
To this end, for example, Treedoc \cite{Preguica09Commutative} uses a tree, where it is always 
possible to insert an element between two other elements when considering an infix 
traversal -- the unique identifier is constructed by combining the path on the tree 
with a pair (timestamp, replica identifier) that makes it unique.

We note that the designs presented for the add-wins set can be used as the basis
for creating a list CRDT. 
For example, for using Treedoc unique identifiers, only the following changes would 
be necessary.
First, replace the function that creates a new unique identifier by the Treedoc function 
that creates a new unique identifier.
We note that in the state-based design that includes a summary of unique identifiers 
observed, the pair (timestamp, replica identifier) is what needs to be recorded.
Second, create a function that returns the data ordered by the unique identifier.

%%%possible to implement every CRDT
%%%just record operations and use a function to compute
%%%
%%%
%%%operations that commute
%%%
%%%count operation based
%%%
%%%
%%%state-based
%%%=> need to divide the space
%%%associative
%%
%%
%%total order
%
%unique identifiers
%lamport clocks
%
%
%non-existent -> created -> removed -> discarded
%
%causal consistency enforces this, so done
%
%
%for state based, necessary to compare a version where there is one uid
%created and removed
%
%summary of unique identifiers
%know whether it is in or not
\subsubsection{Space complexity of CRDTs}
The space complexity of a CRDT depends on the data stored in each moment, 
on the data structures used and on the metadata stored. 
Some old CRDT designs (mostly for state-based synchronization) stored tombstones 
for removed data, which would make the amount of data stored to be larger 
(and sometimes much larger) than the actual data. Modern designs avoid
this overhead, but still typically store metadata that grows linearly 
with the number of replicas  for tracking concurrency and causal 
predecessors \cite{DBLP:journals/ipl/Charron-Bost91}.
For example, Algorithm \ref{alg:st:addwins} shows a state-based add-wins
set CRDT that includes a vector clock with complexity $O(n)$, 
with $n$ the number of replicas.

In this example, for each element in the set, at least one unique identifier is
present.
Although these unique identifiers do no influence the space complexity of the 
CRDTs (as they are of constant size), they represent an important overhead 
for the state of the object (storage and communications). 
To reduce the size of this metadata, possible solutions include 
more compact causality representations when multiple replicas are 
synchronized among the same nodes 
\cite{DBLP:journals/dc/MalkhiT07,swiftcloud,DBLP:conf/srds/GoncalvesABF17,DBLP:journals/corr/AlmeidaB13}. 

A complementary approach proposed by Baquero et. al. \cite{Baquero14Making} 
is to discard meta-data after the relevant updates become stable.
Although the proposal has only been used in the context of pure-operation based 
synchronization and required close integration with the underlying synchronization
subsystem, we believe the idea could be adapted to other synchronization models
by relying on mechanisms
for establishing the stability of updates in weakly consistent systems \cite{Golding92Weak}.

\subsubsection{Time complexity of operations}
The time complexity of operations depends on the data structures used, as
with any implementation of an ADT. We note that the CRDT designs presented in
literature (and in this document) typically do not focus on this issue.
For example, although in the add-wins set design presented in 
Algorithm \ref{alg:st:addwins} the information about elements is maintained in a set, 
for having $O(1)$ expected running time for lookup (add and remove), the actual CRDT 
implementation would use an hash table.

The time complexity of operations in a CRDT are also affected by the need to access and
update the metadata stored in the CRDT.
For example, the merge of the vector clocks in the merge function of 
Algorithm \ref{alg:st:addwins} has complexity $O(n)$, with $n$ the number of 
replicas.

\subsection{Advanced topics}

In this section we discuss other topics that may be relevant for the CRDT developer.

\subsubsection{Reversible computation}
Non trivial Internet services require the composition of multiple sub-systems, 
to provide storage, data dissemination, event notification, monitoring and 
other needed components. 
When composing sub-systems, that can fail independently or simply reject some 
operations, it is useful to provide a CRDT interface that undoes 
previously accepted operations. 
Another scenario that would benefit from undo is collaborative editing of 
shared documents, where undo is typically a feature available to users.

Undoing an increment on a counter CRDT can be achieved by a decrement. 
Logoot-Undo \cite{Weiss10LogootUndo} proposes a solution for undoing (and redoing) 
operations for a sequence CRDT used for collaborative editing.
The implementation of SwiftCloud \cite{swiftcloud} includes
CRDT designs that allow accessing an old value of the CRDT.   
However providing an uniform approach to undoing, reversing, operations over 
the whole CRDT catalog is still an open research direction. 
The support of undo is also likely to limit the level of compression that 
can be applied to CRDT metadata.

\subsubsection{Security}
While access to a CRDT based interface can be restricted at the system level by 
adding authentication and access control mechanisms, 
any accessing replica has the potential to issue operations that can interfere 
with the other replicas. 
For instance, delete operations can remove all existing state. 
In state based CRDTs, replicas have access to state that holds a 
compressed representation of past operations and metadata. 
By manipulation of this state and synchronizing to other replicas, 
it is possible to introduce significant attacks to the system operation 
and even its future evolution. 

Applications that store state on third party entities, such as in 
cloud storage providers, might elect not to trust the provider and 
choose end-to-ends encryption of the exchanged state and use
multiple cloud providers for storing data \cite{depsky}. 
This, however, would require all processing to be done at the edge, 
under the application control. 
A research direction would be to allow some limited form of computation, 
such as merging state, over information whose content is subject to encryption. 
Potential techniques, such as homomorphic encryption, are likely to 
pose significant computational costs. 
An alternative is to execute operations in encrypted data without disclosing it, 
relying on specific hardware support, such as Intel SGX and ARM TrustZone.

\subsubsection{Verification}

An important aspect related with the development of distributed systems that 
use CRDTs is the verification of the correctness of the system. 
This involves not only verifying the correctness of CRDT designs, but also 
the correctness of the system that uses CRDTs. A number of works have addressed 
these issues. 

Regarding the verification of the correctness of CRDTs, several approaches have been 
taken. The most commonly used approach is to have proofs when designs are proposed
or to use some verification tools for the specific data type, such as 
TLA \cite{Lamport94TLA} or Isabelle\footnote{Isabelle: \url{http://isabelle.in.tum.de/}.}.
There has also been some works that proposed general techniques 
for the verification of CRDTs \cite{Burckhardt14Replicated,Zeller14Formal,Gomes17Verifying}, 
which can be used by CRDT developers to verify the correctness of their designs.
Some of these works \cite{Zeller14Formal,Gomes17Verifying} 
include specific frameworks that help the developer 
in the verification process.

A number of other works have proposed techniques to verify the correctness
of distributed systems that run under weak consistency, identifying when
coordination is necessary 
\cite{Li14Automating,Balegas15Putting,Homeostasis,Gotsman16Cause,Zeller17Testing,blazes}.
Some of these works focused on systems that use CRDTs.
Sieve \cite{Li14Automating} computes, for each operation, the weakest precondition 
that guarantees that application invariants hold. 
In runtime, depending on the operation parameters, the system runs an operation under
weak or strong consistency.

Some works \cite{Balegas15Putting,Gotsman16Cause,Zeller17Testing} require the developer 
to specify the properties that the distributed system must maintain, and a 
specification of the operations in the system (that is often independent 
of the actual code of the system). 
As a result, they state which operations cannot execute concurrently.

%When comparing these work, we can 
%identify two slightly different approaches. 
%Indigo \cite{Balegas15Putting} assumes 
%that it is possible to merge any concurrent updates and only verifies
%if the merged state violates application invariants.
%CISE analysis \cite{Gotsman16Cause} is slightly more restrictive, in which
%it verifies if the execution of one operation makes the preconditions
%of other operations invalid. Thus, it only allows the concurrent execution of
%operation that could have been executed sequentially.
%For example, consider a multi-player game where a team of players must have a 
%leader. Each player can assign herself as leader if there is no leader.
%Given these conditions, CISE would require coordination for the update
%that assigns a new leader. 
%With Indigo, if LWW is used for solving concurrent assignments for the leader, 
%there is no need for coordination as the merged state obeys the invariants
%that there are at most one leader.
%Repliss \cite{Zeller17Testing} is the more general tool, allowing to adopt
%both approach, by specifying invariants over the state and over state transitions.

Despite these works, the verification of the correctness of CRDT designs and
of systems that use CRDTs, how these verification techniques can be made available
to programmers, and how to verify the correctness of implementations, remain 
an open research problem.

\section{Final remarks}

CRDTs have been used in a large number of distributed systems and applications
that adopt weak consistency models. 
This document presented an overview of Conflict-free Replicated Data Types research 
and practice, and it was organized in three parts.

The first part discussed the aspects that concern mostly the application developer, 
who uses CRDTs to maintain the data of her application. We argue that the most
important aspect is that of concurrency semantics, which defines the behavior of
the CRDT in the presence of concurrent updates.

The second part discussed the aspects that concern mostly the system developer, 
who is designing a system that embeds CRDTs. We argue that in this case, the
most important aspect is the synchronization model, that defined which properties
the system must enforce to guarantee that CRDTs behave correctly.

Finally, the last part discussed the aspects that concern mostly the CRDT developer,
who want to design new CRDTs. In this case, we identified and explained how a 
number of techniques have been used to design CRDTs.

\section*{Acknowledgments}

We would like to thank Carlos Baquero, Annette Bieniusa, Marc Shapiro, Marek Zawirski, 
J. Legatheaux Martins and Margarida Mamede for their comments and suggestions.
This work was partially supported by NOVA LINCS (UID/CEC/ 04516/2013) and 
EU H2020 LightKone project (732505).

This document was produced as part of the documentation for obtaining the 
\emph{agregação} degree at Universidade NOVA de Lisboa, and it is a modified and extended 
version of Preguiça et. al. \cite{crdt:enc}.

\bibliographystyle{plainurl}
\bibliography{crdts}

\end{document}